\documentclass[12pt]{article}
\usepackage{epsfig}

\def\beq{\begin{equation}}
\def\eeq#1{\label{#1}\end{equation}}
\def\eeqn{\end{equation}}

\def\beqa{\begin{eqnarray}}
\def\eeqa#1{\label{#1}\end{eqnarray}}
\def\eeqan{\end{eqnarray}}

\let\bar=\overbar

\def\Dslash{\not{\hbox{\kern-4pt $D$}}}
\def\dslash{\not{\hbox{\kern-2pt $\del$}}}

\def\msb{{\bar{\ssstyle M \kern -1pt S}}}

\def\Title#1{\begin{center} {\Large {\bf #1} } \end{center}}

\begin{document}

\Title{MuLan Measurement of the Positive Muon Lifetime 
and Determination of the Fermi Constant}

\begin{center}
{\small Proceedings of CKM 2012, the 7th International Workshop 
on the CKM Unitarity \\ Triangle, University of Cincinnati,  USA, 
28 September - 2 October 2012}
\end{center}

\bigskip\bigskip

%+\addtocontents{toc}{{\it D. Reggiano}}
%+\label{ReggianoStart}

\begin{raggedright}  

{\it Tim Gorringe for the MuLan Collaboration, \\
Department of Physics and Astronomy, University of Kentucky, \\
Lexington, KY, 40506, USA}
\bigskip\bigskip
\end{raggedright}

\begin{abstract}
We report results from the MuLan measurement 
of the positive muon lifetime. The experiment was conducted at the Paul Scherrer Institute 
using a time-structured surface muon beam and a segmented plastic scintillator array. 
Two different in-vacuum muon stopping targets were used: a ferromagnetic foil 
with a large internal magnetic field and a quartz crystal in a moderate 
external magnetic field. From a total of $1.6 \times 10^{12}$ decays, 
we obtained the muon lifetime $\tau_{\mu} = 2\, 196\, 980.3(2.2)$~ps (1.0~ppm)
and Fermi constant $G_F = 1.166\, 378\, 7(6) \times 10^{-5}$ GeV$^{-2}$ (0.5~ppm).
\end{abstract}

%\section{Introduction}

The Fermi constant $G_F$ is best determined by measurement 
of the positive muon lifetime $\tau_{\mu}$. 
Following work  by  van Ritbergen and Stuart \cite{vanRitbergen:1999fi}
and Pak and Czarnecki \cite{Pak:2008qt}---that
reduced the theoretical uncertainty in the relation 
between the muon lifetime and the Fermi constant to 0.14~pm---two new 
experiments have been performed to measure $\tau_{\mu}$ \cite{Webber:2010zf, Barczyk:2007hp}.
Herein we report the results from the part-per-million
measurement of the positive muon lifetime by the MuLan Collaboration.

\section{Experimental setup}

The experiment was conducted at the Paul Scherrer Institute 
using a nearly 100\% longitudinally polarized, 29~MeV/c, $\mu^+$ beam 
from the $\pi$E3 secondary beamline
at the 590~MeV proton cyclotron. 
%Surface muons originate from  at-rest $\pi^+ \rightarrow \mu^+ \nu_{\mu}$ decay 
%in the outer layer of the production target and yield an intense source 
%of $\sim$100\% longitudinally polarized, 29~MeV/c positive muons.
Incoming muons were stopped in solid targets and outgoing positrons
were detected in a finely segmented, fast timing, scintillator array.
The photomultiplier signals from scintillator detectors were recorded
by 450~MHz sampling-rate, 8-bit resolution, waveform digitizers
and read out by a high-speed data acquisition.

One important feature of the experiment was the application 
of a time structure on the muon beam.
The time structure consisted of a 5~$\mu$s-long, 
beam-on ``accumulation period'' followed by
a 22~$\mu$s-long, beam-off ``measurement period''.
The structure was imposed on the continuous beam using a
custom-built, fast-switching, 25~kV electrostatic kicker.
With the kicker high voltage off, the muons were transported 
to the target. With the kicker high voltage 
on, the muons were deflected into a collimator.
%(the beam extinction during measuring periods was roughly 900).
The structure was important in avoiding the 
need to associate positrons with parent muons, a requirement that limited the
statistics of earlier measurements.

Another important feature of the experiment was the setup and
the choice of the muon stopping targets.
First, the $\pi$E3 beamline was extended through the positron detector
with  the stopping target mounted in the beam vacuum.
This design reduced the number of upstream stops.
Second, two combinations of stopping target materials and transverse magnetic fields 
were used: a magnetized Fe-Cr-Co foil (Arnokrome-III) with a 
4~kG internal $B$-field and a quartz crystal disk (SiO$_2$)
in a 130~G external $B$-field. 
In the ferromagnetic target, where muons reside as diamagnetic ions, 
the $\mu^+$ precession frequency  was $\sim$50~MHz.
In the quartz target, where muons reside mostly as paramagnetic atoms,
the $\mu^+e^-$ precession frequency was $\sim$180~MHz.
The fast precession yielded a roughly 1000-fold
reduction in the ensemble-averaged $\mu^+$ polarization 
via the spin dephasing during the beam accumulation.

\begin{figure}
\begin{center}
\epsfig{file=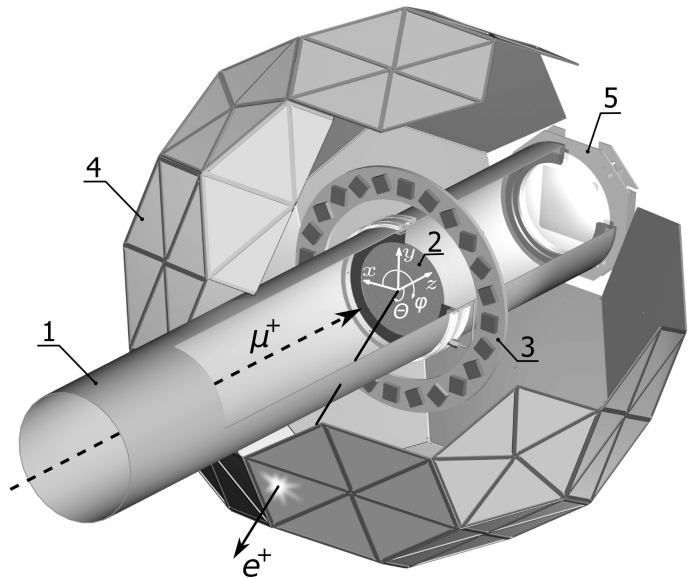,height=3.0in}
\caption{Cutaway drawing of the experimental setup: (1) vacuum beampipe, (2) stopping target,
(3) halbach magnet, (4) scintillator array, and (5) beam monitor.}
\label{fig:detector}
\end{center}
\end{figure}

The positron detector was constructed of 170 triangle-shaped plastic scintillator pairs  
arranged in a soccer ball (truncated icosahedral) geometry (Fig.~\ref{fig:detector}).
Each pair comprised an inner plastic scintillator tile 
and outer plastic scintillator tile. The pairs were grouped into ten
pentagonal enclosures containing five tile-pairs 
and twenty hexagonal enclosures containing 
six tile-pairs, which together formed the soccer ball geometry. 
The segmentation was important in reducing 
positron pile-up in individual detector elements.
The symmetric arrangement of detector elements
was important in reducing the effects of muon spin rotation / relaxation ($\mu$SR).

% detector
%\begin{figure}
%\begin{center}
%\psfig{file=detector.ps}
%\end{center}
%\caption{Cutaway drawing of the experimental setup showing the vacuum beampipe (1), 
%stopping target (2), Halbach magnet (3), scintillator array (4)
%and the beam monitor (5) at the downstream window of the beam line.
%Note that the Halbach magnet was not installed for the AK-3 data taking
%and the stopping target was rotated out of the
%beam path for the beam monitoring.}
%\label{fig:MuLan_ball}
%\end{figure}

\section{Data analysis}

A total of 1.0$\times$10$^{12}$ decays from stops in
Arnokrome-III  were collected  in our 2006 production run
and 0.7$\times$10$^{12}$ decays from stops in
quartz were collected  in our 2007 production run.
The runs yielded about 130~terabytes of digitized pulses.
%For consistency checks and systematic studies,
%the data were accumulated with different orientations of the magnetic field 
%and different centering of the muon stopping distribution in the positron detector. 
The analysis was conducted using the 88-teraflop ABE cluster at the National Center 
for Supercomputing Applications (NCSA). 
%The processing of digitized waveforms 
%into time histograms took roughly 60,000 CPU-hours for each dataset.

The time and amplitude of pulses were determined 
from least square fits to the individual digitized waveforms.
The procedure involved fitting a 
relatively high-resolution standard waveform (0.022~ns sampling-interval) 
to a relatively low-resolution individual waveform (2.2~ns sampling-interval). 
The high-resolution standard waveforms
were constructed by combining a large sample 
of 450~MHz sampling-rate, single-pulse, digitized waveforms. 
If the quality of the fit to a single pulse was not acceptable,
the algorithm attempted to improve the fit by adding pulses or removing pulses.

Two cuts were applied on the amplitudes and the times of the
hits before constructing inner-outer tile-pair coincidences.
One cut defined an unambiguous software amplitude 
threshold $A_{thr}$ for detector hits. Another cut defined 
an unambiguous artificial deadtime $ADT$
between detector hits. Hits that survived these cuts
were sorted into time distributions 
of inner singles, outer singles and inner-outer coincidences.
The construction of coincidence histograms with different thresholds $A_{thr}$
and artificial deadtimes $ADT$ was important for studying such effects 
as pulse pileup and gain variations.

\subsection{Systematics effects}

%A part-per-million measurement of the muon lifetime requires a careful analysis
%of many systematic effects. Most important, were any distortions of time distributions
%from  positron pulse pileup and gain variaitions.
% pileup correction
If a hit occurs in the artificial deadtime of an earlier hit it is lost. 
Because these losses are more frequent during the high rates at early measurement times
than the low rates at late measurement times, such pile-up distorts the time spectrum.

Our procedure for correcting for pileup took advantage of the time structure 
of the incident beam. The pileup losses were statistically recovered 
by replacing the lost hits in each measurement period with
measured hits at equivalent times in neighboring measurement periods.
For example, to correct for leading-order pileup, 
if a hit is observed at time $t_i$ in fill $j$ (the ``trigger'' hit),
a hit is searched for within the interval $t_i$~$\rightarrow$~$t_i + \mathrm{ADT}$ 
in fill $j+1$ (the ``shadow'' hit).  Adding the resulting histogram 
of shadow hit times to the original histogram of trigger hit times
thereby statistically recovers the lost hits.
Similar methods were used to correct for higher-order pileup and
accidental coincidences.

% gain correction
In filling histograms only hits with amplitudes that exceed 
the threshold $A_{thr}$ were used.
Consequently, if the detector gain varies over the
 measurement period,  then the time histogram will be distorted,
either by additional hits rising above the amplitude cut or  
by additional hits falling below the amplitude cut.

Our procedure for correcting for gain variations took advantage of the measurement
of the positron pulse amplitude during the measurement period.
The detector gain versus measurement time 
was determined from the MIP (minimally ionizing particle) peak 
of the decay positrons in the scintillator tiles.
Using the MIP-peak position versus measurement time, 
the time histograms were corrected for gain variations.

\subsection{Lifetime fits}

%ak3 method
A simple procedure was used to extract the lifetime $\tau_{\mu}$ 
from the Arnokrome-III target.
The summed tile-pair time histogram of coincidence hits 
was fit to  $N e^{-t/\tau_{\mu}} + C$.
The approach relied on sufficient cancellation 
of Arnokrome-III $\mu$SR effects by spin dephasing
and detector geometry. No evidence of $\mu$SR distortion
were observed in the time spectra from the Arnokrome-III target.

A more detailed procedure was needed to extract the lifetime $\tau_{\mu}$ 
from the quartz target.
First, geometry-dependent effective lifetimes were extracted
for each tile-pair from fits to
\begin{equation}
N(t) = N e^{-t/\tau_\mathrm{eff}} [ 1 + f(t) ] + C ,
\label{eq:Lambda:d}
\end{equation}
where $f(t)$ accounts for time-dependent effects of transverse-field (TF) spin precession / relaxation.
Then, the lifetime $\tau_{\mu}$ is extracted from the 170 effective lifetimes $\tau_\mathrm{eff}$
from a fit to 
\begin{equation}
\tau_\mathrm{eff} ( \theta_B , \phi_B  ) = \tau_{\mu} ( 1 +  \delta ( \theta_B , \phi_B ) )  ,
\label{eq:fitfunction:sophisticated}
\end{equation}
where $\delta ( \theta_B , \phi_B )$
accounts for geometry-dependent effects of longitudinal-field (LF) spin relaxation.
Together these steps accounted for all observed features 
of quartz TF/LF $\mu$SR.

Note a number of datasets were accumulated with different
magnetic field orientations and different positron detector offsets.
These datasets changed the magnitude and orientation of
TF/LF $\mu$SR effects. The agreement between lifetime results 
for different configurations was an important verification
of the fitting procedures.

\section{Results}
%results

The individual results \cite{Webber:2010zf} for the muon lifetime 
from the Arnokrome-III dataset and the quartz dataset
are 
\begin{equation}
\tau_{\mu}({\rm Arnokrome-III}) =  2~196~979.9 \pm 2.5 (stat) \pm 0.9 (syst) {\rm ~ps}
\end{equation}
and
\begin{equation}
\tau_{\mu}({\rm quartz}) = 2~196~981.2 \pm 3.7 (stat) \pm 0.9 (syst) {\rm ~ps}.
\end{equation}
The combined result 
\begin{equation}
\tau_{\mu}({\rm MuLan}) = 2~196~980.3 \pm  2.1 (stat) \pm 0.7 (syst) {\rm ~ps}
\label{finalresult}
\end{equation}
is obtained from the weighted average of the individual  values with
the appropriate accounting for the correlated uncertainties.
The result corresponds to an overall uncertainty in the muon lifetime 
of 2.2~ps or 1.0~ppm---a thirty-fold improvement over earlier generations
of lifetime measurements. The largest contributions to systematic uncertainties 
are associated with the aforementioned pulse pileup, gain changes
and $\mu$SR effects, as well as the knowledge of the time independence of the 
beam extinction during the measurement period.

The final result for $\tau_{\mu}$ is in reasonable agreement 
with the earlier work of Duclos {\it et al.}, Balandin {\it et al.},
Giovanetti {\it et al.} and Bardin {\it et al.} 
However, it is in  marginal disagreement (2.9~$\sigma$) with the 
commissioning result from the FAST collaboration experiment \cite{Barczyk:2007hp}.

We use the relation obtained by van Ritbergen and Stuart (vRS) 
\cite{vanRitbergen:1999fi}
for the determination of the Fermi constant $G_F$ from the
measurement of the muon lifetime $\tau_{\mu}$. 
The vRS relation was derived using the
$V$-$A$ current-current Fermi interaction with 
QED corrections evaluated to 2-loop order. It yields
$G_F ({\rm MuLan}) =  1.166\, 378\, 7(6)\times 10^{-5}{GeV^{-2}} (0.5~\mathrm{ppm})$, 
a thirty-fold improvement over the 1998 Particle Data Group \cite{PDG1998}
value that pre-dates the vRS theoretical work
and recent lifetime measurements.
The 0.5~ppm error is dominated by the 1.0~ppm uncertainty of the lifetime
measurement, with contributions of 0.08~ppm from 
the muon mass measurement and 0.14~ppm from the theoretical corrections.

\end{document}